\documentclass[11pt,twoside]{article}

\usepackage{asp2004}
\usepackage{graphicx}

\begin{document}

\title{ High resolution CO imaging of high redshift QSO host galaxies}  
\author{F. Walter$^{1}$, D.~A. Riechers$^{1}$, C.~L. Carilli$^{2}$, F. 
Bertoldi$^{3}$, A. Weiss$^{4}$ and P. Cox$^{5}$}   
\affil{$^{1}$Max-Planck-Institut f\"ur Astronomie, K\"onigstuhl 17, 
Heidelberg, D-69117, Germany\\
  $^{2}$National Radio Astronomy Observatory, PO Box O, Socorro, NM 
87801, USA\\
  $^{3}$Argelander-Institut f\"ur Astronomie, Auf dem H\"ugel 71, Bonn, 
D-53121, Germany\\
  $^{4}$Max-Planck-Institut f\"ur Radioastronomie, Auf dem H\"ugel 69, 
Bonn, D-53121, Germany\\
  $^{5}$ Institut de RadioAstronomie Millim\'etrique, 300 Rue de la 
Piscine,
Domaine Universitaire, 38406 Saint Martin d'H\'eres, France}

\begin{abstract} 
  We review recent high-resolution CO observations of distant QSOs
  obtained at the Very Large Array.  The aim of these observations is
  to resolve the molecular gas distribution in these extreme objects
  both spatially and in velocity space. They provide unique
  information regarding the small-scale distribution, the extent, and
  the brightness temperatures of the molecular gas in these early
  systems. E.g., the structure and dynamics of the molecular gas may
  reveal whether or not mergers can be the cause of the ongoing
  starburst activity. The observations also allow for a first estimate
  of the dynamical gas mass. Currently, only the VLA is able to obtain
  resolutions in CO of up to 0.15$''$ which is needed to resolve
  typical galactic structures of sizes $\sim$1 kpc. We present new
  high-resolution VLA imaging of high--$z$ QSOs (BRI 1335--0417, APM
  08279+5255 and J1148+5251).  These observations pave the road to
  future ALMA observations where resolutions of order 0.1$''$ will be
  obtained routinely.
\end{abstract}



\vspace{-1cm}

\section{Introduction: QSOs at high $z$}

Over the last few years, the study of high redshift QSOs has been
revolutionized in three ways. First, wide field surveys have revealed
100's of high--$z$ QSOs, right back to the epoch of cosmic reionization
($z > 6$; e.g., Fan et al. 2006). Second, it has been shown that most
(all?) low redshift spheroidal galaxies have central super-massive
black holes (SMBH), and that the black hole mass correlates with bulge
velocity dispersion.  This M$_{BH}$--$\sigma_{\rm v}$ correlation
suggests coeval formation of galaxies and SMBH, thereby making SMBHs a
fundamental aspect of the galaxy formation process (Gebhardt et al.
2000).  And third, mm surveys of high redshift QSOs find that 30$\%$
of the sources are `hyper-luminous infrared galaxies' ($L_{FIR} =
10^{13}$ L$_\odot$), corresponding to thermal emission from warm dust,
and that this fraction is {\sl independent of redshift out to $z =
  6.4$} (Beelen et al. 2004, Omont et al. 2004). If the dust is heated
by star formation, the implied star formation rates are extreme ($>
10^3$ M$_\odot$ year$^{-1}$), consistent with the formation of a large
elliptical galaxy on a dynamical timescale of 10$^8$ years.  \\

\section{Molecular Gas in QSO hosts}

Molecular line observations (typically CO) of FIR-luminous high--$z$
QSOs have revealed large gas masses in most cases observed to date
(see Carilli et al.\ 2004; Solomon \& Vanden Bout 2005). Detecting
large amounts of warm, extended molecular gas currently provides the
strongest evidence that these luminous QSOs are undergoing vigorous
star formation. The coeval growth of massive black holes and of
massive stellar populations can be examined directly in these unique
systems, providing an opportunity to study the cause of the tight
correlation between these components that is observed in local
spheroidal galaxies.  High resolution observations are crucial for
this, since the structure of the dense gas which is feeding the
starburst and the black hole can reveal the cause of event (e.g.
mergers) and the mass of the systems, placing them into the cosmic
structure evolution context. Molecular gas has now been detected in
more than a dozen $z>2$ QSO host galaxies.  Most sources have been
studied in the higher order transitions ($\ge$ CO 3--2), although at
$z \ge 4$ the lower order transitions become accessible to cm
telescopes such as the VLA.

The {\em detection} of CO emission is critical to estimate the
reservoir of the molecular gas in these early systems. A second step
is then to {\em spatially resolve the molecular gas distribution}. In
particular, given the typical diameters of galaxies of many kpc, a
linear resolution of $\sim 1$\,kpc is needed to {\em resolve the
  structure} of the underlying host galaxy. Such measurements are
needed 1) to get an estimate for the {\em size of the host galaxy}
(and thus a better estimate for the dynamical mass), 2) to resolve
potentially merging systems and 3) to better constrain the {\em
  physical properties} of the molecular gas by measuring the
brightness temperature of the molecular gas in the hosts. A linear
resolution of 1\,kpc corresponds to a resolution of 0.15$''$ at the
redshifts under consideration (1$''\sim8.5$\,kpc at $z=2$,
1$''\sim5.8$\,kpc at $z=6$). Such observations can then in turn be
used to constrain the predictions by CDM simulations of early galaxy
formation, and, if a large sample was available, put limits on the
frequency of mergers at high redshift. In addition, such studies can
be used to constrain the possible redshift-evolution of the M$_{\rm
  BH}$--$\sigma_{\rm v}$ relation (as will be discussed below).

\begin{figure}[h]
\hspace{3cm}
\includegraphics[width=7.5cm]{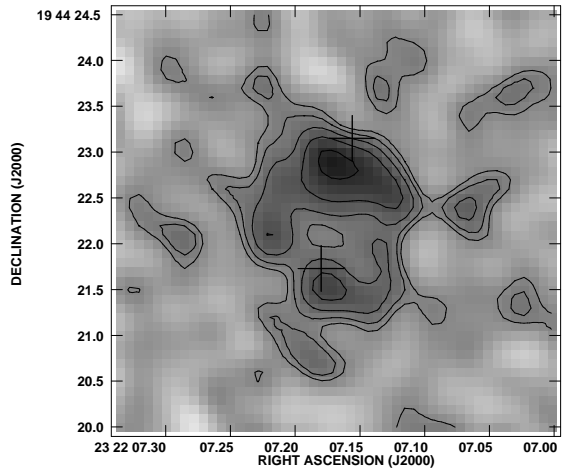}
\vspace*{-10mm}

\caption{Image of the CO(2--1) emission from PSS J2322+1944 (taken from 
Carilli et al.\ 2003) at 0.6$''$ resolution.  The crosses show the 
positions of the optical QSOs, and the cross sizes represent the 
relative astrometric error.  The contour levels are a geometric 
progression in square root 2 starting at 0.12 mJy beam$^{-1}$.}
\end{figure}

It is important to note that the highest resolution imaging of QSO
host galaxies can currently only be obtained using the recently
upgraded Q--band system at the VLA. In optical/NIR observations the
central bright AGN greatly overshines the host galaxy, thus rendering
optical/NIR studies of the QSO hosts impossible given current
instrumentations. We note that observations of the molecular gas phase
at resolutions of 0.15$''$ will even be difficult to achieve with ALMA
during the 'early science' operations, as they require $\sim$4\,km
($\nu\sim100$\,GHz) or $\sim2$\,km ($\nu\sim200$\,GHz) baselines
(numbers are for a $z\sim2.5$ source).  For comparison, current mm
interferometers typically only observe at baselines up to
$\sim400$\,m.

\begin{figure}[h]
\includegraphics[width=12cm]{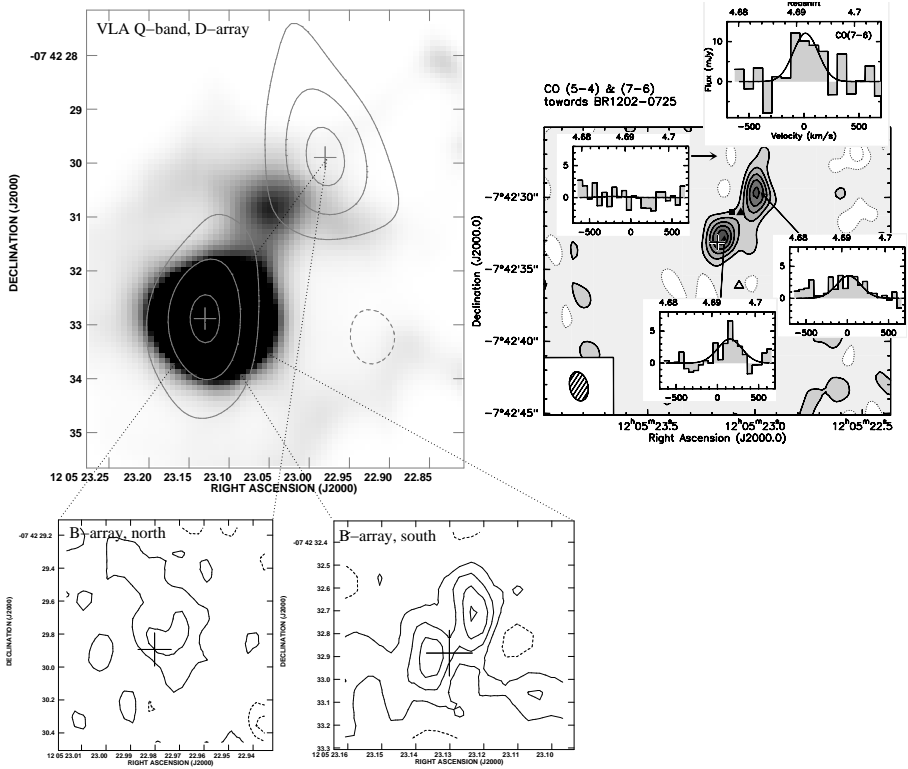}
\vspace*{-5mm}

\caption{{\em left}: CO(2--1) imaging of BR\,1202-0725 using the VLA 
D--array
  (Carilli et al.\ 2002).  Contours are at -0.13,0.13,0.26,
  0.39\,mJy/beam. The greyscale is a narrow-band image of the
  Ly--$\alpha$ emission.  {\em bottom}: VLA B--array {\em only} data
  of the northern source ({\em b.left panel}) and the southern source
  ({\em b.right panel}).  Contours are shown at -0.2, -0.1, 0.1, 0.2,
  0.3\,mJy/beam.  {\em right}: Plateau de Bure spectra of the CO(5--4)
  and CO(7--6) line emission ($\sim$3.5$''$ resolution) superimposed to the
  1.35\,mm continuum image (Omont et al.\ 1996).  In both the PdBI
  and the VLA D--array observations, there are two clearly separated
  components. }
\end{figure}

\section{Previous Work}

Over the last few years, the VLA has been used to image a number of
high--$z$ sources. In the following, we will briefly review some of
these observations and their implications for understanding the
properties of molecular gas at the highest redshifts.

{\bf J2322+1944 ($z=4.1$): } In Fig.~1 we show the CO emission from
the $z = 4.12$ QSO J2322+1944. The source is strongly gravitationally
lensed, appearing as a double QSO in the optical with 1$''$ separation
(Djorgovski et al.\ in prep), and a complete 'Einstein ring' in the CO
2-1 emission (Carilli et al.\ 2003). This molecular ring makes
J2322+1944 one of the most stunning CO images at high redshift. Quite
importantly, a similar ring is seen for the 1.4 GHz continuum emission
(Carilli et al.\ 2003). These observations have been modeled (by the
above mentioned authors) in the source plane as a starburst disk
surrounding the QSO with a radius of about 2 kpc.  J2322+1944 thus
represents perhaps {\em the best example of a coeval starburst+AGN} at
high redshift. This exemplifies that strongly lensed sources can be
used to even study the overall morphologies in galaxies (if a good
lens model exists).

\begin{figure}
\includegraphics[width=13cm]{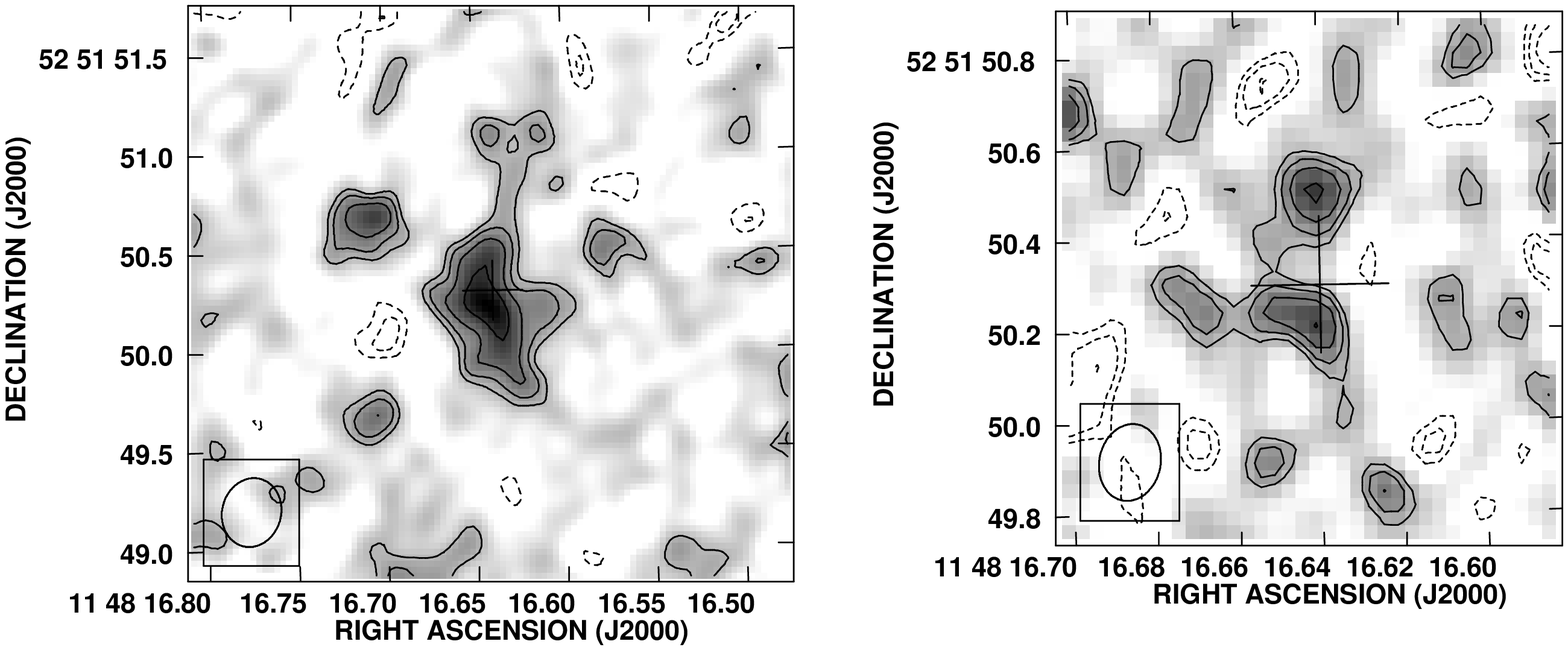}
\vspace*{-5mm}

\caption{ {\em left}:
  CO(3--2) map of J1148+5251 of the combined B- and C-array data sets
  (covering the total bandwidth, 37.5 MHz or 240 km s$^{-1}$). Contours are
  shown at -2, -1.4, 1.4, 2, 2.8, and 4 $\times \sigma$ (1$\sigma$ = 43 
$\mu$Jy beam$^{-1}$). The beam size (0.35$''$ $\times$ 0.30$''$) is shown in 
the bottom left
  corner.  {\em right}: 
  The central region at 1\,kpc (0.17$''$ $\times$ 0.13$''$) resolution 
(1$\sigma$ = 45 $\mu$Jy beam$^{-1}$).
2, 2.8, and 4 $\times \sigma$ (1$\sigma$ = 45 $\mu$Jy beam$^{-1}$).
  This high--resolution map
  recovers half of the flux seen in the left panel (see Walter et al.\
  2004 for more details). New HST ACS imaging suggests that the AGN
  seen in the optical is associated with the souther CO clump seen in
  the right panel (White et al.\ 2005).}
\end{figure}

{\bf B1202-0725 ($z=4.7$): } At $z = 4.7$, BR\,1202-0725 is observed at
a cosmic time of $\sim$1.3\,Gyr after the Big Bang. It was detected in
multiple CO transitions and recently also in CO(1--0) using the GBT
(see Riechers et al. 2006, also this volume).  This source has the
curious property that the optical QSO is a single source, but the mm
continuum and CO line observations show a double source with a
separation of about 4$''$ (Omont et al.\ 1996; Guilloteau et al.\
1999; Carilli et al.\ 2002). This is illustrated in the CO imaging
shown in Fig.~2. This double morphology may indicate a pair of
interacting objects separated by about only 20\,kpc (Yun et al.\ 2000)
or a QSO exhibiting a massive starburst (southern source) with a
dust--obscured, Ly--$\alpha$ - emitting companion (northern source, Hu
et al.\ 1996) ionized by the strong QSO, which also contains a
starburst.  The total far-IR luminosity of the ULIRG BR\,1202-0725 is
$L_{\rm FIR} = 4.2 \times 10^{13}\,L_{\odot}$, leading to a $L_{\rm
  FIR}/L'_{\rm CO(1-0)}$ ratio of 350. The implied total molecular gas
mass in this system is about $5 \times 10^{10}\,M_{\odot}$ (Carilli et
al.\ 2002).

{\bf J1148+5251 ($z=6.4$): } This source is still the redshift
record--holder for QSOs. It has been detected in CO(3-2) using the VLA
(Walter et al.\ 2003) and in CO(6--5) and CO(7--6) using the Plateau
de Bure interferomter (Bertoldi et al.\ 2003). Follow--up VLA B and C
array observations clearly resolved the the molecular distribution
spatially and also in velocity space (Walter et al.\ 2004, see
Fig.~3). The molecular gas distribution in J1148+5251 is extended out
to radii of 2.5\,kpc. The central region is resolved and shows 2
peaks, separated by 1.7\,kpc; they account for about half of the total
emission, with the other half present in the more extended molecular
gas distribution. Each of these peaks harbors a molecular gas mass of
$\sim5\times10^9$\,M$_{\odot}$ within a radius of 0.5\,kpc,
respectively; this mass is similar to the total mass found in nearby
ULIRGS such as Mrk\,273 or Arp\,220 (Downes \& Solomon 1998). The
peaks have intrinsic brightnesses of $\sim35$\,K (averaged over the
1\,kpc-sized beam), similar to what is found in the centres of nearby
active galaxies (Downes \& Solomon 1998, assuming constant surface
brightness down to the CO(1--0) transition), albeit measured over a
larger physical area.

Based on the extent of the molecular gas distribution and the
line-width measured from the higher CO transitions we derive a
dynamical mass of $\sim4.5\times10^{10}$\,M$_{\odot}$
($\sim5.5\times10^{10}$\,M$_{\odot}$ if we correct for an inclination
of $i\sim65^{\circ}$).  This dynamical mass estimate can account for
the detected molecular gas mass within this radius but leaves little
room for other matter. In particular, given a black hole mass of mass
$\sim$1-5$\times10^9$ M$_{\odot}$ (Willot et al.\ 2003), this
dynamical mass could not accomodate an order few$\,\times
10^{12}$\,M$_{\odot}$ stellar bulge which is predicted by the
present--day M$_{\rm BH}-\sigma_{\rm bulge}$ relation (Ferrarese \&
Merritt 2000, Gebhardt 2000), if this relation were to hold at these
high redshifts. Even if we assume a scenario in which this bulge was
10 times the scale length of the molecular gas emission, we would
still expect a bulge contribution of few$\,\times
10^{11}$\,M$_{\odot}$ within the central 2.5\,kpc (assuming a central
density--profile of $\rho\sim r^{-2}$, e.g., Jaffe 1983, Tremaine
1994) which can not be reconciled with our results. Our finding
therefore suggests that black holes may assemble before the stellar
bulges. Depending on the space density of similar sources at these
high redshifts, the smaller dynamical masses may be in better
agreement with the masses predicted by CDM simulations in the very
early universe as a 10$^{12}$\,M$_{\odot}$ bulge would imply a rather
massive dark matter halo of $>10^{14}$\,M$_{\odot}$. Regardless, these
results show the power of high-resolution CO imaging in sources at the
end of cosmic reionization.

\begin{figure}[h]
\vspace{0cm}
\hspace{3cm}
\includegraphics[width=8cm]{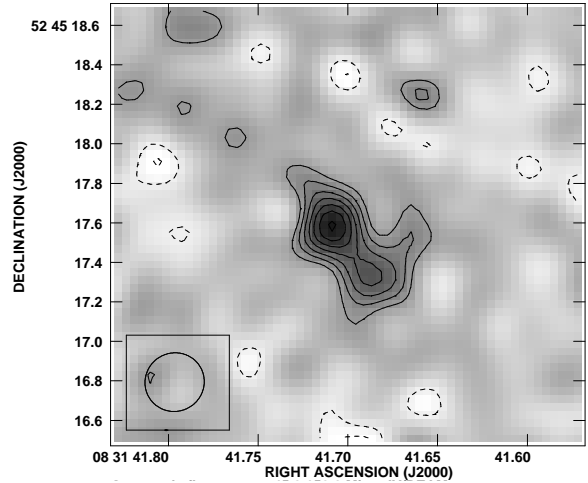}
\vspace*{-1mm}

\caption{Image of CO(1--0) towards APM\,08729+5255 obtained with the VLA
  covering a bandwidth of 43.75\,MHz, or 560\,km s$^{-1}$ (Riechers et
  al.\ 2006, in prep.).  Contours are shown at -3, -2, 2, 3, 4, 5, 6,
  7, and 8 $\times \sigma$ (1$\sigma$ = 16 $\mu$Jy beam$^{-1}$. The beam
  size (0.30$''$ $\times$ 0.30$''$) is shown in the bottom left
  corner. We find no evidence for an extended reservoir of molecular
  gas around this source.}

\end{figure}

\section{New Observations}

In the following, we present some new observations that we recently
obtained at the VLA in B array configuration (D. Riechers, PhD Thesis,
MPIA Heidelberg).

{\bf APM0827+5255 ($z=3.9$): } The presence of extended CO(1--0)
emission in APM0827+5255 was reported by Papadopoulus et al.\ (2001).
Additional CO(2--1) C--array imaging by Papadopoulus et al.\ also
showed some evidence for an extended envelope around the central QSO.
We have have obtained new C and B array observations for this source
and do not find evidence for such an extended CO component. In Fig.~4,
we show our highest resolution data, where we detect a compact but
resolved structure of the lens images. Two clearly separated images
can be seen, and their morphology and brightess ratio is very similar
to that of the optical QSO.  We thus conclude that the CO in this
source is indeed extremely compact, and that the lensing magnification
is close to the optical magnification i.e., a factor of 100 rather
than 7 as derived in previous models (Lewis et al.\ 2002). Further
evidence against an extended CO(1--0) reservoir around APM0827+5255
comes from our multi--CO transition study of this source (Weiss et
al.\ 2006, in prep., see also this volume, and recent single dish
observations of this source using the GBT, Riechers et al.\ 2006, see
also this volume).

\begin{figure}[h]
\vspace{0cm}
\hspace{3cm}
\includegraphics[width=7.8cm]{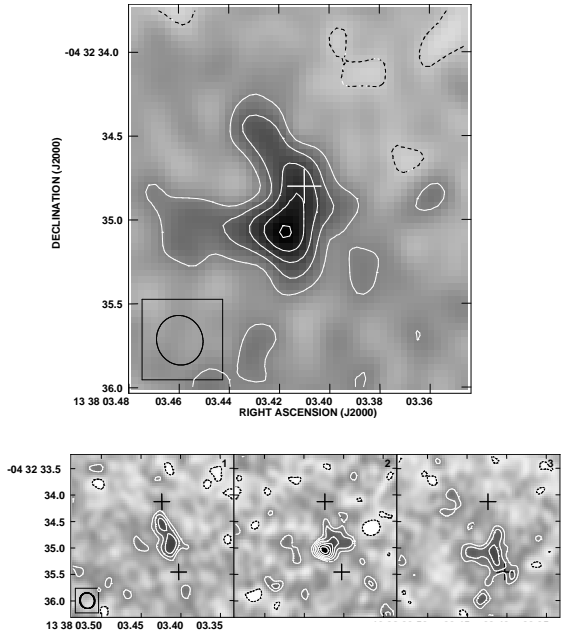}

\caption{Image of CO(2--1) towards BRI\,1335-0417 obtained with the VLA.
  {\em top}: Moment 0 map covering a bandwidth of 43.75\,MHz, or
  300\,km s$^{-1}$ (Riechers et al.\ 2006, in prep.).  Contours are
  shown at -2, 2, 4, 6, 8, and 10 $\times \sigma$ (1$\sigma$ = 50 $\mu$Jy
  beam$^{-1}$.  {\em bottom}: Channel maps with a width of 12.5\,MHz,
  or 85\,km s$^{-1}$ each, covering the central 260 \,km s$^{-1}$ of
  the emission.  Contours are shown at -3, -2, 2, 3, 4, 5, 6, 7, and 8
  $\times \sigma$ (1$\sigma$ = 100 $\mu$Jy beam$^{-1}$). For the first time
  these observations allow us to study the kinematics in a high-z QSO
  host galaxy in detail. }
\end{figure}

{\bf B1335-0417 ($z=4.4$): } This is one of the first sources detected
in CO(2--1) with the VLA in D array (Carilli et al.\ 1999). We
recently re--observed the CO(2--1) transition in the $z=4.4$ QSO
BRI\,1335-0417 at high resolution (0.15$''$) with the VLA in B--array
(Fig.\ 5, Riechers et al.\ 2006, in prep.). The integrated CO map
shows a compact but resolved molecular gas reservoir, and our velocity
channel maps show, for the first time, the actual dynamics of the
molecular gas in a high-z QSO host galaxy at high signal--to--noise.
The detected structure shows an interesting morphology which may be
interpreted as a close merger or rotation. We are currently modeling
the kinematics of this unique dataset.

\section{Concluding remarks}

The recent observations presented here have shown that resolved
molecular gas {\em maps} of objects out to the highest redshifts can
already be achieved with today's instrument (i.e. the VLA). Although
these observations are time intensive, they have dramatically improved
our knowledge of the properties of the molecular gas phase at high
redshift. E.g., they have helped to constrain the source sizes and
kinematics, which in turn constrain the dynamical masses (e.g. in the
case of J1148+5251, and B1335-0417). Also, they have provided clear
evidence for co-eval starburst and AGN activity (e.g., J2322+1944) and
revealed likely interactions with objects that are barely visible in
the optical (e.g.  B1202-0725). These results show the enormous
potential for future studies of the molecular gas content and
dynamical masses in a statistically significant sample of
high--redshift galaxies using ALMA, where resolutions of $<0.1''$ will
be achieved routinely.

\acknowledgements 
D.~A.\ R.\ acknowledges support from the Deutsche Forschungsgemeinschaft
(DFG) Priority Programme 1177.
C.~C.\ would like to acknowledge support from the
Max-Planck-Forschungspreis.




\begin{thebibliography}{}
\bibitem[2004]{bee04}
Beelen, A., et al.\ 2004, A\&A, 423, 441
\bibitem[2003]{ber03}
Bertoldi, F., Cox, P., Neri, R., et al.\ 2003, A\&A, 409, L47
\bibitem[1999]{car99}
Carilli, C.~L., Menten, K.~M., \& Yun, M.~S.\ 1999, 521, L25
\bibitem[2002]{car02}
Carilli, C.~L., Kohno, K., Kawabe, R., et al.\ 2002b, AJ, 123, 1838
\bibitem[2003]{car03}
Carilli, C.~L., Lewis, G.~F., Djorgovski, S.~G., et al.\ 2003, Science, 
300, 773
\bibitem[2004]{car04}
Carilli, C., et al. 2004, astro-ph/0402573
\bibitem[1998]{dow98}
Downes, D., \& Solomon, P.~M. 1998, ApJ, 507, 615
\bibitem[2006]{fan06}
Fan, X, et al.\ 2006, AJ, 131, 1203
\bibitem[2000]{fm2000}
Ferrarese, L.~\& Merritt, D.\ 2000, \apjl, 539, L9
\bibitem[2000]{g2000}
Gebhardt, K., et al.\ 2000, \apjl, 539, L13
\bibitem[1999]{gui99}
Guilloteau, S., et al.\ 1999, A\&A 349, 363
\bibitem[1996]{hu96}
Hu, E., et al.\ 1996, ApJ, 459, L53
\bibitem[1983]{jaf83}
Jaffe, W.\ 1983, \mnras, 202, 995
\bibitem[2002]{lew02}
Lewis, G.~F., Carilli, C., Papadopoulos, P., \& Ivison, R.~J. 2002,
MNRAS, 330, L15
\bibitem[1996]{omo96}
Omont, A., Petitjean, P., Guilloteau, S., et al.\ 1996, Nature, 382, 428
\bibitem[2004]{omo04}
Omont, A., Beelen, A., Bertoldi, F., Carilli, C.~L., \& Cox, P.\ 2004, in
'Multiwavelength AGN Surveys',World Scientific, Singapore, 2004, 109
\bibitem[2001]{pap01}
Papadopoulos, P., Ivison, R., Carilli, C.~L., \& Lewis, G. 2001, Nature, 
409, 58
\bibitem[2006]{rie06}
Riechers, D.~A., Walter, F., Carilli, C.~L., et al.\ 2006, ApJ, in press
\bibitem[2005]{sv05}
Solomon, P.~M., \& Vanden Bout, P.~A. 2005, ARA\&A, 43, 677
\bibitem[1994]{tre94}
Tremaine, S., Richstone, D.~O., Byun, Y., et al.\ 1994, \aj, 107, 634
614, L97
\bibitem[2003]{wal03}
Walter, F., Bertoldi, F., Carilli, C.~L., et al.\ 2003, Nature, 424, 406
\bibitem[2004]{wal04}
Walter, F., Carilli, C., Bertoldi, F., et al.\ 2004, ApJ, 615, L17
365, 271
\bibitem[2005]{whi05}
White, R. L, Becker, R. H, Fan, X., \& Strauss, M. A., 2005, AJ, 129, 2102
\bibitem[2003]{w03}
Willott, C.~J., McLure, R.~J., \& Jarvis, M.~J.\ 2003, \apjl, 587, L15
\bibitem[2000]{yun00}
Yun, M., et al.\ 2000, ApJ, 528, 171

\end{thebibliography}
\end{document}